\documentclass[aps,pra,10pt,superscriptaddress,twocolumn,floatfix,nofootinbib]{revtex4-2}
\usepackage{graphicx}
\usepackage{amsmath,amsthm,amssymb,dsfont,soul,enumitem}
\usepackage{xcolor}
\usepackage{mdframed}
\usepackage{comment}
\usepackage{orcidlink}
\usepackage{placeins}
\usepackage{float}
\usepackage{tikz}
\usepackage[normalem]{ulem}
\usepackage[commandnameprefix=ifneeded,authormarkuptext=name,commentmarkup=uwave]{changes}
\usetikzlibrary{arrows.meta,positioning,calc,shapes.misc}
\newcommand{\ket}[1]{\left| #1 \right\rangle}
\newcommand{\bra}[1]{\left\langle #1 \right|}

\newcommand{\ketbra}[2]{\left|#1\right\rangle\hskip-1mm\left\langle#2\right|}

\definechangesauthor[name={EB},color=green!70!black]{Enrico}

\usepackage{hyperref}
\usepackage[capitalise]{cleveref}
\crefname{equation}{Eq.}{Eqs.}
\crefname{figure}{Fig.}{Figs.}
\crefname{section}{Sec.}{Secs.}
\crefname{table}{Table}{Tables}
\crefname{theorem}{Theorem}{Theorems}
\crefname{lemma}{Lemma}{Lemmas}
\crefname{proposition}{Proposition}{Propositions}
\crefname{corollary}{Corollary}{Corollaries}
\crefname{definition}{Definition}{Definitions}
\crefname{appendix}{Appendix}{Appendices}

\theoremstyle{definition}

\newtheorem{proposition}{Proposition}

\newcommand{\id}{{\mathds{1}}}
\newcommand{\tr}{\mathrm{Tr}}

\newcommand{\Ee}{\mathcal{E}}

\newcommand{\Z}{\mathbb{Z}}
\newcommand{\Mana}{\mathcal{M}}
\newcommand{\sn}{\mathrm{sn}}
\newcommand{\CF}{\mathrm{CF}}
\newcommand{\E}{\mathbb{E}}

\soulregister{\cite}{7}\soulregister{\cref}{7}\soulregister{\Cref}{7}\soulregister{\ref}{7}\soulregister{\eqref}{7}\soulregister{\emph}{1}\soulregister{\textit}{1}

\newmdenv[backgroundcolor=yellow!50,linewidth=0pt,innerleftmargin=3pt,innerrightmargin=3pt,innertopmargin=3pt,innerbottommargin=3pt,skipabove=6pt,skipbelow=6pt]{addedblock}

\begin{document}

\title{Transition in Contextuality for random quantum circuits}
\author{Enrico Bozzetto}
\affiliation{DET, Politecnico di Torino, Corso Duca degli Abruzzi, 24, 10129 Torino, Italy}
\affiliation{Quantum Group, School of Computing, Newcastle University, 1 Science Square, Newcastle upon Tyne, NE4 5TG, UK}
\author{Jonte R. Hance\,\orcidlink{0000-0001-8587-7618}}
\email{jonte.hance@newcastle.ac.uk}
\affiliation{Quantum Group, School of Computing, Newcastle University, 1 Science Square, Newcastle upon Tyne, NE4 5TG, UK}

\begin{abstract}
Niroula \textit{et al.}~\cite{Niroula_2024} showed that a random stabiliser code subject to coherent errors exhibits a phase transition in magic: coherent errors create magic, and compete with syndrome measurements, which can destroy magic. We ask whether the same transition can be seen in contextuality.  In the qutrit version of the model, Kochen--Specker contextuality with respect to stabiliser measurements vanishes on the same logical states as Wigner negativity, and we find numerically the same behaviour as for magic (up to $N=12$ qutrits for the mana and $N=8$ for the contextual fraction). Unlike magic, Spekkens contextuality of one logical qubit, tested with stabiliser preparations and measurements, vanishes on all $T$-type magic states~\cite{BozzettoHance2026ClassicalLimit}, yet its circuit average peaks where magic does. For two or more logical qutrits it again vanishes on the same states as Wigner negativity, with a noise robustness set by the most negative Wigner value. 
We also run the qubit model on the five-qubit IQM processor of Politecnico di Torino and recover the peak at about $60\%$ of its noiseless height.

\end{abstract}
\maketitle
\section{Introduction}
Niroula \textit{et al.}~\cite{Niroula_2024} showed that magic undergoes a phase transition in a simple and experimentally accessible model: a random Clifford code subject to coherent single-qubit rotations of angle $\alpha$. Syndrome measurements, which can destroy magic, compete with the errors, which create it. With a single logical qubit, the logical state has finite magic with a peak at a distance of order $1/\sqrt{N}$ from the Clifford point $\alpha=\pi/2$. At constant rate $r=K/N$ this critical region becomes an extended magical phase. The fitted critical point and exponent differ between the diagnostics used (second stabiliser R\'enyi entropy~(SSRE)~\cite{Leone2022SRE} or conditional measurement entropy).

Later work identified the mechanism. Encoding--decoding circuits with coherent errors have an error-resilience transition, beyond which the decoder no longer returns the encoded state~\cite{Turkeshi2024}. With the syndrome fixed or post-selected, the magic transition coincides with it~\cite{Sierant2026}. With syndromes sampled by the Born rule, as in Ref.~\cite{Niroula_2024} and here, the syndrome probabilities act as a relevant perturbation: they shift the critical point, change the exponent and cause strong finite-size drifts~\cite{Sierant2026}. Thresholds in random stabiliser codes have also been located through the recoverability of logical information, for errors given by random Clifford unitaries applied with probability $p$~\cite{Liu2026Coherent}. Under amplitude damping, with post-selection on the trivial syndrome, the decoding fidelity still has a transition but the robustness of magic does not~\cite{BallarTrigueros2026}. Magic transitions also occur in doped Clifford circuits~\cite{Leone2024Transition,Magni2025}, monitored Clifford+$T$ circuits~\cite{Bejan2024}, hybrid circuits~\cite{Fux2024} and measurement-only circuits~\cite{Tarabunga2025}, and coherent errors followed by syndrome measurement can generate logical unitary designs~\cite{Cheng2025}. In a fixed code with a unitary error on a correctable region that supports no stabiliser, syndrome extraction removes the robustness of magic that the error adds and returns that of the logical state~\cite{BozzettoHance2026RoM}.

Contextuality is the other standard way to formalise ``non-classicality as a computational resource''. For qudits of odd prime dimension, contextuality with respect to stabiliser measurements is necessary for universal computation by magic-state distillation~\cite{Howard2014} and is equivalent to negativity of the discrete Wigner function~\cite{Howard2014,Delfosse2017} (for one qudit with an ancilla~\cite{Howard2014}, and for $n\ge2$ qudits of any odd dimension~\cite{Delfosse2017}). Wigner negativity is itself necessary for magic-state distillation~\cite{Veitch2012}.

This suggests that the magic transition of Ref.~\cite{Niroula_2024} should have a contextuality counterpart. In this work we study it.

First, the main difference between contextuality and magic is that magic is a property of the state itself, while contextuality is a property of a scenario, i.e. a set of states (or a single state) and a set of measurements (a set of transformations can be added if wanted). Second, for two or more qubits the stabiliser states are themselves contextual with respect to the Pauli observables~\cite{Peres1990,Mermin1990,Mermin1993}, so there is no ``noncontextual at the Clifford points'' baseline analogous to ``zero magic at the Clifford points''~\cite{Niroula_2024}. Third, contextuality is a binary property so a quantifier is needed to see a transition at all.

Our work considers four different approaches.
\begin{enumerate}
\item  We take the Niroula model to qutrits, where Wigner negativity is equivalent to contextuality with respect to stabiliser measurements on $n\ge2$ qutrits, and study the density of mana of the logical state in the vanishing-rate and constant-rate regimes.
\item We quantify Kochen--Specker (KS) contextuality~\cite{KochenSpecker1967} of two logical qutrits by the contextual fraction, computed exactly by a linear program with a true zero baseline.
\item We quantify Spekkens contextuality of one logical qubit and one logical qutrit by the robustness of simplex-embeddability, computed exactly by a linear program; for one qubit it has a closed form~\cite{BozzettoHance2026ClassicalLimit}, and for two or more qutrits we show that it is a function of the minimum of the Wigner function.
\item We estimate Spekkens contextuality in the one logical qubit scenario on real hardware using the five-qubit IQM Spark quantum computer of Politecnico di Torino.
\end{enumerate}

Our results are of two kinds. Where contextuality is equivalent to Wigner negativity, as for KS contextuality of $n\ge2$ qutrits with respect to stabiliser measurements, its quantifiers vanish on the same logical states as mana (\cref{prop:zeroset}, a consequence of Refs.~\cite{Delfosse2017,Gross2006}), so contextuality and magic vanish at the same error angles, and our numerics give how large the contextual fraction is elsewhere. For Spekkens contextuality, which we test in prepare-and-measure scenarios (the stabiliser states and the logical state, measured in the stabiliser bases), the two notions separate for one logical qubit or qutrit, and coincide again for two or more logical qutrits, where the noise robustness depends only on the minimum of the Wigner function (\cref{prop:multi}). Along the circuits this robustness follows the magic peak at vanishing rate, but at constant rate it approaches $1$ at all non-Clifford angles, so it does not show the constant-rate transition (\cref{fig:spk-rate}). For one logical qubit the noise robustness of Spekkens contextuality has a closed form, which vanishes on all $T$-type magic states~\cite{BozzettoHance2026ClassicalLimit}. We show that the pure states on which it vanishes cover $31\%$ of the Bloch sphere and that it never exceeds $1-1/\sqrt2$ (\cref{prop:qubit}). For one logical qutrit the strange state $(\ket1-\ket2)/\sqrt2$, which has maximal mana~\cite{Veitch2014,Jain_2020}, also has a noncontextual model in this fragment~\cite{BozzettoHance2026ClassicalLimit} (\cref{prop:strange}). If the coherent error is itself part of the scenario, as a transformation, the results of Refs.~\cite{Schmid2022StabSub,BozzettoHance2026ClassicalLimit} exclude a convex-linear, diagram-preserving noncontextual model at every non-Clifford angle, so there is no transition (\cref{sec:discussion}).

Zhang and Gu~\cite{Zhang2024Magic} study magic dynamics in random circuits of odd-prime-dimensional qudits through Wigner negativity and mana, using a statistical-mechanics mapping with a replica trick; their circuits (Haar brickwork circuits, or Clifford brickwork with magic injection by single-qudit Haar-random unitaries) and their analytical regime (large local dimension) differ from the coherent-error code considered here. Their quantities are the ones we use numerically, and their method is a candidate for the analytics we leave open.

\section{setup}
\subsection{Reference model: qubits platform}
\label{sec:setup-qubit}
We begin from the same model used in Ref.~\cite{Niroula_2024} valid for qubits. The initial state is a product state of $N$ qubits in the $\ket{0}$ state, so $\ket{0}^{\otimes N}$. We follow then the same steps followed in Ref.~\cite{Niroula_2024}
\begin{enumerate}
\item \emph{Encoding.} A random Clifford circuit $C$ is applied. In Ref.~\cite{Niroula_2024}, $C$ is built from $d$ layers of single-qubit Cliffords, sampled uniformly from the 24-element single-qubit Clifford group, interleaved with $d$ layers of fixed-angle M{\o}lmer--S{\o}rensen gates $\mathrm{MS}(\pi/2)=e^{i\pi\sigma_x\sigma_x/4}$ on $N/2$ randomly chosen disjoint pairs, with $d=N$ in the numerics.
\item \emph{Error.} A single-qubit rotation $R_z(\alpha)=\exp(-i\alpha\sigma_z/2)$ is applied to each qubit; $\alpha$ is the error rate (an angle). The Clifford points are $\alpha\in(\pi/2)\mathbb{Z}$.
\item \emph{Decoding.} The conjugate circuit $C^\dagger$ is applied.
\item \emph{Syndrome.} The last $N-K$ qubits are measured in the computational basis. These outcomes $s$ are the syndrome of the Clifford code, and they leave a pure state of $K$ logical qubits.
\end{enumerate}
The code rate is $r=K/N$; the vanishing-rate case is $K=1$.

If we denote as $\rho_{C,s}(\alpha)$ the post-syndrome logical state, magic can be evaluated on $\rho_{C,s}(\alpha)$ directly, while contextuality cannot, because it requires a measurement scenario. Having specified that scenario, the qubit setup still presents two problems in terms of our scope.

The first fact is that for more than one qubit, also stabiliser states are contextual. To show that we consider the full Pauli scenario on $K=2$ qubits: the 15 non-identity Pauli observables, with the 15 maximal contexts of three mutually commuting observables. A noncontextual deterministic assignment is a map from the 15 observables to $\pm1$ respecting the product relation inside every context. We enumerated all $2^{15}$ maps and found that none is consistent with all 15 relations (this is the Peres--Mermin obstruction~\cite{Peres1990,Mermin1990,Mermin1993}). Since every quantum state is supported only on outcomes that obey the product relations, no state admits a noncontextual model~\cite{BermejoVega2017}, and the contextual fraction of every state, stabiliser states included, is $1$. For $K\ge2$ the Pauli scenario therefore cannot define an order parameter that vanishes at the Clifford points. 

The second fact is that if we consider only one qubit we cannot exhibit KS contextuality since the Pauli observables $X,Y,Z$ on one qubit are pairwise incompatible, so each observable forms its own context and every empirical model is noncontextual (it is the product of its marginals).

For these reasons for Wigner negativity and KS contextuality we will consider the same platform but based on qutrits. For Spekkens contextuality instead we will use both qutrits and qubits.

\subsection{Qutrits platform}\label{sec:qutrit}
Let $d=3$ and $\omega=e^{2\pi i/3}$, with $X\ket{j}=\ket{j+1}$, $Z\ket{j}=\omega^{j}\ket{j}$, the Fourier gate $F\ket{j}=3^{-1/2}\sum_k\omega^{jk}\ket{k}$, and the phase gate $S=\mathrm{diag}(1,1,\omega)=\mathrm{diag}(\omega^{2^{-1}j(j-1)})$, where $2^{-1}$ is the inverse of $2$ modulo $3$. Conjugation by $S$ maps $X\mapsto XZ$ (up to a phase), so $S$ is Clifford.

The model we use has the same structure as the one described in \cref{sec:setup-qubit}. We consider $N$ qutrits in $\ket{0}^{\otimes N}$, a random Clifford encoder $C$, a product error layer $U(\alpha)^{\otimes N}$, the decoder $C^\dagger$, and computational-basis measurement of the last $N-K$ qutrits giving the syndrome $s\in\Z_3^{N-K}$. The error in this new setup is a phase on one level
\begin{equation}
U(\alpha)=\ket{0}\!\bra{0}+\ket{1}\!\bra{1}+e^{i\alpha}\ket{2}\!\bra{2}
\end{equation}
which is the direct analogue of $R_z(\alpha)\propto\mathrm{diag}(1,e^{i\alpha})$. Since every function $\Z_3\to\Z_3$ is a polynomial of degree at most two, $U(2\pi/3)=S$ and $U(4\pi/3)=S^2$ are Clifford, so the Clifford points are $\alpha\in(2\pi/3)\mathbb{Z}$, as $\alpha\in(\pi/2)\mathbb{Z}$ for qubits. The single-qutrit Clifford group (up to phases) is generated by $F,S,X,Z$ and has $|\mathrm{SL}(2,\Z_3)|\cdot 3^2=24\cdot9=216$ elements. The first term is the number of $2\times 2$ matrices with entries in $\mathbb{Z}_3$ and determinant equal to 1 modulo 3. These are the matrices which transform the Pauli exponents linearly. The second term is the number of possible arrangements of the Pauli exponents. The encoder has $d=N$ layers of single-qutrit gates, each drawn uniformly from these 216 elements and applied independently to every qutrit, interleaved with $d=N$ layers of the two-qutrit gate $\mathrm{CSUM}\ket{a,b}=\ket{a,a+b}$ applied to $\lfloor N/2\rfloor$ random disjoint ordered pairs. The decoder is the inverse circuit. The logical qutrits are the first $K$.

Given $C$, $\alpha$ and a syndrome $s$, the logical state is
\begin{equation}
\ket{\psi_{C,s}(\alpha)}=\frac{(\bra{s}\otimes\id_K)\,C^\dagger U(\alpha)^{\otimes N}C\ket{0}^{\otimes N}}{\sqrt{p_C(s;\alpha)}},
\end{equation}
with $p_C(s;\alpha)$ the probability of the syndrome. For a quantity $Q$ of the logical state we write the circuit- and syndrome-averaged value as
\begin{equation}
\overline{Q}(\alpha;N,K)=\E_C\sum_{s}p_C(s;\alpha)\,Q\bigl(\psi_{C,s}(\alpha)\bigr).
\end{equation}

\section{Identifying contextuality}
We use three notions in the following: Wigner negativity described by mana and sum negativity, KS contextuality, for which we use the contextual fraction, and Spekkens contextuality, quantified through robustness of simplex embeddability. In fragments like ours, which contain complete eigenbases of every measurement mixing to $\id/D$, with $\id/D$ in the interior of the hull of the stabiliser states, Spekkens noncontextuality implies KS noncontextuality for the same measurements (see the proof of \cref{prop:multi} and Refs.~\cite{Spekkens2005,Schmid2022StabSub}; Ref.~\cite{BozzettoHance2026Warring} compares the two notions more generally), but not conversely: one qubit has no KS contextuality, yet the fragments of \cref{prop:qubit} can be Spekkens contextual.

\subsection{Wigner negativity}
For $n$ qudits of odd prime dimension $d$ we use the discrete Wigner function~\cite{Gross2006}. For a pure state $\ket\psi$, with $q,p\in\Z_d^n$,
\begin{equation}
W_\psi(q,p)=d^{-n}\sum_{x\in\Z_d^n}\omega^{2\,p\cdot x}\,\overline{\psi(q+x)}\,\psi(q-x),\quad \sum_{q,p}W_\psi=1
\end{equation}
Two standard quantifiers of its negativity are the sum negativity and the mana~\cite{Veitch2014} (we use the natural logarithm throughout):
\begin{equation}
\sn(\rho)=\frac{\|W_\rho\|-1}{2},\qquad \Mana(\rho)=\log\|W_\rho\| 
\end{equation}

Stabiliser states have $W\ge0$ and $\Mana=0$, and the single-qutrit magic state $(\ket1-\ket2)/\sqrt2$, the strange state~\cite{Veitch2014}, for example has $W_{\min}=-1/3$ and $\Mana=\log(5/3)$~\cite{Jain_2020}.

We use the following facts.
\begin{enumerate}
\item A pure state is a stabiliser state if and only if its Wigner function is non-negative (discrete Hudson theorem~\cite{Gross2006}). Among pure states, $\sn=\Mana=0$ exactly on stabiliser states; mixed states with $W\ge0$ need not be mixtures of stabiliser states~\cite{Gross2006}.
\item Clifford unitaries act on $W_\rho$ by affine symplectic maps of phase space, so $\sn$ and $\Mana$ are Clifford invariant.
\item $\Mana$ is non-increasing under stabiliser operations, including computational-basis measurements, on average over outcomes~\cite{Veitch2014}. This is the counterpart of the fact that syndrome measurements can destroy magic of Ref.~\cite{Niroula_2024}.
\item For odd local dimension $d>1$ and $n\ge2$ qudits, a state $\rho$ (pure or mixed) admits a noncontextual set of deterministic value assignments for stabiliser (Heisenberg--Weyl) measurements if and only if $W_\rho\ge0$~\cite{Delfosse2017}. The proposition there (Prop.~1 of Ref.~\cite{Delfosse2017}) requires $n\ge2$. For a single qudit of odd prime dimension, Howard \textit{et al.}~\cite{Howard2014} show that states with negative Wigner function are contextual with respect to stabiliser measurements, but their construction uses two-qudit stabiliser projectors: the single-qudit stabiliser projectors alone detect nothing, since each belongs to exactly one context. As every context is then a single basis, any single-qudit state admits a noncontextual model for these measurements (the product of its marginals).
\end{enumerate}

We assess a single logical qutrit through $\rho\otimes\ket0\!\bra0$, on which $n=2$ stabiliser measurements are defined. Since $W_{\rho\otimes\sigma}=W_\rho\otimes W_\sigma$ and $\ket0$ has $W\ge0$ with $\|W\|=1$, we have $\Mana(\rho\otimes\ket0\!\bra0)=\Mana(\rho)$, and $\rho\otimes\ket0\!\bra0$ is contextual if and only if $W_\rho$ has a negative entry.

To be able to see the transition we will use the densities $\overline{\Mana}/K$ and $\overline{\sn}/K$ as order parameters, as Ref.~\cite{Niroula_2024} uses magic densities. For a fixed encoder, $\sum_s p_C(s;\alpha)\,\sn(\psi_{C,s})$ is the sum negativity of the code-adapted Wigner representation of $U(\alpha)^{\otimes N}C\ket0^{\otimes N}$ (Theorem~IV.1 of Ref.~\cite{BozzettoHance2026PhaseSpace}), so $\overline{\sn}$ is also a property of the state before the syndrome measurement.

\subsection{KS contextuality: the contextual fraction}
\label{sec:ks}
To quantify KS contextuality we use the contextual fraction~\cite{Abramsky2011,Abramsky2017CF}. For an empirical model $e$ on a scenario with contexts $C$, the contextual fraction 
\begin{equation}
\CF(e)=1-\max\{\lambda: e=\lambda e^{\mathrm{NC}}+(1-\lambda)e'\}
\end{equation}
with $e^{\mathrm{NC}}$ noncontextual and $e^\prime$ any no-signalling model. It vanishes if and only if $e$ is noncontextual, and is $1$ if and only if $e$ is strongly contextual~\cite{Abramsky2011}. For qubits the full Pauli scenario on $K\ge2$ gives $\CF=1$. 

We begin considering the two qutrit scenario, which can then be generalised to $n>2 $ qutrits. 
The observables are all the 80 non-identity Heisenberg-Weyl operators on two qutrits and the contexts are the 40 maximal commuting sets. The states are the 360 stabiliser states $\sigma$, whose Wigner function equals 1/9 on an affine Lagrangian plane $P_\sigma$ of nine phase-space points. The deterministic noncontextual assignments are the $81$ phase-space points~\cite{Delfosse2017}; the point $x$ assigns outcome $\sigma$ iff $x\in P_\sigma$. The probability of the event is $\tr(\sigma\rho)=\sum_{x\in P_\sigma}W_\rho(x)$. The contextual fraction is therefore
\begin{equation}
\CF(\rho)=1-\max_{w\ge0}\Bigl\{\sum_x w_x:\ \sum_{x\in P}w_x\le\sum_{x\in P}W_\rho(x)\ \ \forall P\Bigr\},
\label{eq:cf}
\end{equation}
a linear program in $81$ variables with $360$ constraints.

The same setup can be generalised to more than two qutrits. However the number of variables of the linear program grows as $9^K$ and the constraints as the number of affine Lagrangian planes, $3^K\prod_{k=1}^K(3^k+1)$: $729$ variables and $30\,240$ constraints for $K=3$, and $6561$ variables and about $7.4\times10^6$ constraints for $K=4$.

\begin{proposition}\label{prop:zeroset}
Let $\psi$ be a pure state of $n\ge2$ qutrits and let $\CF(\psi)$ be its contextual fraction for the Heisenberg--Weyl scenario of \cref{eq:cf}, generalised to $n$ qutrits. Then $\CF(\psi)=0$ if and only if $\sn(\psi)=\Mana(\psi)=0$, if and only if $\psi$ is a stabiliser state, and $\CF$, $\sn$ and $\Mana$ are invariant under Clifford unitaries. In the model of \cref{sec:qutrit} (using $\rho\otimes\ket0\!\bra0$ for $K=1$) every logical state is a stabiliser state at the Clifford points $\alpha\in(2\pi/3)\Z$, and for every $\alpha$, $\overline{\CF}(\alpha)=0$ if and only if $\overline\Mana(\alpha)=0$.
\end{proposition}
\begin{proof}
For $n\ge2$, $\psi$ is noncontextual if and only if $W_\psi\ge0$ (fact 4), and $\sn=\Mana=0$ if and only if $W_\psi\ge0$. For pure states this holds if and only if $\psi$ is a stabiliser state (fact 1). A Clifford unitary acts on the Wigner function by an affine symplectic map~\cite{Gross2006}, which permutes the phase-space points and the affine Lagrangian planes, so it permutes the variables and constraints of \cref{eq:cf} and leaves $\|W\|$ unchanged. At $\alpha\in(2\pi/3)\Z$, $U(\alpha)^{\otimes N}$ is Clifford, so $C^\dagger U(\alpha)^{\otimes N}C\ket0^{\otimes N}$ is a stabiliser state, and so is each logical state: its Wigner function, $p_C(s;\alpha)^{-1}\sum_{p_B}W(u;q_B{=}s,p_B)$ with $u$ a phase-space point of the logical qutrits and $B$ the syndrome qutrits, is non-negative, and it is pure, so it is a stabiliser state by the discrete Hudson theorem~\cite{Gross2006}. Finally, $\overline{\CF}$ and $\overline\Mana$ are averages of non-negative terms with weights $p_C(s;\alpha)$, and the terms vanish for the same logical states.
\end{proof}

\Cref{prop:zeroset} shows that $\overline{\CF}$ vanishes exactly where $\overline\Mana$ does, in particular at the Clifford points, but not how large it is elsewhere. The numerics of \cref{sec:ksresult} supply the latter.

\subsection{Spekkens contextuality}
For Spekkens (generalised) contextuality~\cite{Spekkens2005} we use simplex embeddability~\cite{Schmid2021SimplexEmbed}. A GPT is a quadruple $G=(V,\langle\cdot,\cdot\rangle,\Omega,\Ee)$: $\Omega$ is a compact convex state space, $\Ee\subseteq\Omega^*$ a compact convex effect set containing $0$ and a unit effect $u$ with $\langle u,s\rangle=1$ for all $s\in\Omega$. It is \emph{simplex-embeddable} if there exist an inner product space $W$ containing a simplex $\Delta_D$ and its dual hypercube $\Delta_D^*$, and linear maps $\iota,\kappa:V\to W$ with $\iota(\Omega)\subseteq\Delta_D$, $\kappa(\Ee)\subseteq\Delta_D^*$ and $\langle e,s\rangle_V=\langle\kappa(e),\iota(s)\rangle_W$ for all $e\in\Ee$, $s\in\Omega$. A prepare-measure experiment admits a generalised-noncontextual ontological model with a finite ontic state space if and only if its GPT is simplex-embeddable~\cite{Schmid2021SimplexEmbed}. For finitely many states and effects this can be decided by a linear program~\cite{GittonWoods2022,Selby_2024}. For $K$ logical qudits, take the fragment with the stabiliser states together with $\rho$ as states and the Pauli (stabiliser) measurement effects. For $K=1$ qubit the stabiliser part (the six Pauli eigenstates and the six Pauli effects) is the prepare-and-measure part of Spekkens' toy bit~\cite{Spekkens2007Toy}, which has a noncontextual model with four ontic states, so the baseline is exactly zero. With the Clifford transformations included and represented convex-linearly, the single-qubit stabiliser subtheory is contextual~\cite{Lillystone2019,BozzettoHance2026ClassicalLimit} (Ref.~\cite{KociaLove2019} gives a model on non-disjoint ontic states, outside the class of models this applies to), so this zero baseline holds for the prepare-and-measure fragment only. Since we want to quantify the contextuality to see the transition we use the simplex-embeddability robustness as criterion, i.e.

\begin{equation}
    \begin{aligned}
\varepsilon^*(\rho)=\min\Bigl\{\varepsilon\in[0,1]:\ \mathcal{F}_K\bigl((1-\varepsilon)\rho+\varepsilon\,\id/d^K\bigr)\\\text{is simplex-embeddable}\Bigr\}
\end{aligned}
\end{equation}
where $\mathcal{F}_K$ is the fragment represented by the GPT containing all the $K$-qudit stabiliser states, the additional state $\rho$ and all stabiliser measurements. This criterion answers the question ``How much white noise must be added to the quantum state \(\rho\) before the resulting prepare-and-measure theory admits a generalised noncontextual ontological model?''. A larger value of $\varepsilon^*(\rho)$ means that more noise is required to eliminate the contextuality, so the state is more robustly contextual according to this criterion.

\subsubsection{Selby \textit{et al.}'s linear program}
As an alternative method to quantify Spekkens contextuality we use also the criterion described by Selby \textit{et al.} in Ref.~\cite{Selby_2024}. In this work the authors describe a linear program to test simplex embeddability of a fragment and its robustness, using a cone formulation. A fragment is composed of the spans of states and effects, with the Born rule as a bilinear form $B$. One takes the facets $H_S$ of the state cone (rows) and the facets $H_E$ of the effect cone (columns) and asks whether $B=H_E^{T}\sigma H_S$ with $\sigma \ge 0$. To obtain a quantification of \textit{robustness} of simplex-embeddability (and by consequence of contextuality) the authors introduce depolarising noise, $D_r(s)=(1-r)s+r\langle u,s\rangle m$, with $m$ the maximally mixed state. This allows one to define the \textit{ global robustness} (called the robustness of nonclassicality in Ref.~\cite{Selby_2024}) as 
\begin{equation}
\begin{aligned}r(\mathcal F)=\min\{r\ge0:\ &(1-r)B+r\,B_{\mathrm{dep}}=H_E^{T}\,\sigma\,H_S,\\ &\sigma\ge0\}\end{aligned}
\label{eq:rglobal}
\end{equation}
where $B_{\mathrm{dep}}$ is the Born matrix of $|m\rangle\langle u|$. The robustness is then the least $r$ for which the fragment $\{D_r(s)\}$ is simplex-embeddable. It is a single linear program. The main difference between this code and the one we developed is that in the latter the noise affects only the state, while in Ref.~\cite{Selby_2024}'s one the depolarising noise acts globally. Moreover, unlike $\varepsilon^*$, $r$ is defined also when the stabiliser fragment is not embeddable. We compare $r$ with $\varepsilon^*$ in Appendix~\ref{app:lp}.

\subsubsection{One logical qubit and one logical qutrit}
The first part of \cref{prop:qubit}, \cref{eq:epsqubit} and the formula for $\varepsilon_{\rm stab}$, are in Sec.~X of Ref.~\cite{BozzettoHance2026ClassicalLimit} (Theorem~5 and the text before it), since $\min(m_+,m_-)=\|\mathbf r\|_1-2\min_i|r_i|$. We give a short proof in the present notation. Ref.~\cite{BozzettoHance2026ClassicalLimit} states that the $T$ and $H$ values are the extremes of this formula. The proof that $\varepsilon^*\le1-1/\sqrt2$, with equality only on the $H$-type states, is added here.

\begin{proposition}[One logical qubit]\label{prop:qubit}
Let $T_+=\{(1,1,1),(1,-1,-1),(-1,1,-1),(-1,-1,1)\}$ and, for a Bloch vector $\mathbf r$, let $m_\pm(\mathbf r)=\max_{\mathbf v\in T_+}(\pm\mathbf v\cdot\mathbf r)$. For $K=1$ the fragment made of the six Pauli eigenstates, $\rho$ and the six Pauli effects is simplex-embeddable if and only if $\mathbf r$ lies in one of the two tetrahedra $\mathrm{conv}(T_+)$ and $\mathrm{conv}(-T_+)$, and
\begin{equation}\varepsilon^*(\rho)=\max\bigl\{0,\,1-1/\min(m_+,m_-)\bigr\}.\label{eq:epsqubit}\end{equation}
The white-noise robustness of magic, $\varepsilon_{\rm stab}(\rho)=\min\{\varepsilon:(1-\varepsilon)\rho+\varepsilon\id/2 \text{ is a mixture of stabiliser states}\}$, is $\max\{0,1-1/\max(m_+,m_-)\}=\max\{0,1-1/\|\mathbf r\|_1\}=1-1/\mathcal R(\rho)$, with $\mathcal R$ the robustness of magic~\cite{HowardCampbell2017}, which for one qubit is $\max(1,\|\mathbf r\|_1)$~\cite{BozzettoHance2026ClassicalLimit}. Hence $\varepsilon^*\le\varepsilon_{\rm stab}$. Moreover $\varepsilon^*\le1-1/\sqrt2$, with equality only on the $H$-type states, whose Bloch vectors have two components equal to $\pm1/\sqrt2$ and the third zero~\cite{BravyiKitaev2005}.
\end{proposition}
\begin{proof}
The cone generated by the Pauli effects has the eight facets $(1,\mathbf v)$, $\mathbf v\in\{\pm1\}^3$, the deterministic assignments to $X,Y,Z$. By the criterion of Refs.~\cite{Schmid2021SimplexEmbed,Selby_2024}, the fragment is simplex-embeddable if and only if there are functionals $\mu_{\mathbf v}$, affine in $\mathbf r$ and non-negative on every state of the fragment, with $\sum_{\mathbf v}\mu_{\mathbf v}(\mathbf r)(1,\mathbf v)=(1,\mathbf r)$. Expanding in the functions $1,v_i,v_iv_j,v_1v_2v_3$ on $\{\pm1\}^3$ gives $\mu_{\mathbf v}=\tfrac18[1+\mathbf v\cdot\mathbf r+\sum_a g_a(\mathbf r)\chi_a(\mathbf v)]$ with $\chi_a\in\{v_1v_2,v_1v_3,v_2v_3,v_1v_2v_3\}$ and $g_a$ affine. At $\pm\mathbf e_i$ the four $\mu_{\mathbf v}$ with $1\pm v_i=0$ must vanish, and their correction terms sum to zero; this forces $\mu_{\mathbf v}(\mathbf r)=\tfrac18(1+\mathbf v\cdot\mathbf r)(1+\delta\,v_1v_2v_3)$ with $|\delta|\le1$. For $|\delta|<1$ the state must satisfy $1+\mathbf v\cdot\mathbf r\ge0$ for all eight $\mathbf v$, i.e.\ lie in the stabiliser octahedron. This is the one-parameter family of Ref.~\cite{WallmanBartlett2012} on eight ontic states (the family $x_\lambda$ of Ref.~\cite{BozzettoHance2026ClassicalLimit}). The cases $\delta=\pm1$ are its four-state reductions, $\mu_{\mathbf v}=(1+\mathbf v\cdot\mathbf r)/4$ on $T_+$ or $-T_+$, which are the qubit discrete Wigner functions of the two inequivalent quantum nets of Ref.~\cite{GibbonsHoffmanWootters2004} and give the toy bit of Spekkens~\cite{WallmanBartlett2012,Spekkens2007Toy}, valid on $\mathrm{conv}(\pm T_+)=\{\mathbf r:\mathbf v\cdot\mathbf r\ge-1\ \forall\mathbf v\in\pm T_+\}$, which contain the octahedron. Since $\mathbf r\in\mathrm{conv}(T_+)$ if and only if $m_-\le1$, and $\mathbf r\in\mathrm{conv}(-T_+)$ if and only if $m_+\le1$, white noise $\varepsilon$ is enough if and only if $(1-\varepsilon)\min(m_+,m_-)\le1$, which gives \cref{eq:epsqubit}. The octahedron is $\|\mathbf r\|_1\le1$ and $\|\mathbf r\|_1=\max(m_+,m_-)$. Finally, if $\mathbf v\in T_+$ and $\mathbf w\in-T_+$ attain $m_+$ and $m_-$, then $\min(m_+,m_-)\le(\mathbf v+\mathbf w)\cdot\mathbf r/2\le|\mathbf v+\mathbf w||\mathbf r|/2$, and $\mathbf v,\mathbf w$ differ either in one coordinate, so $|\mathbf v+\mathbf w|=2\sqrt2$, or in all three, so $\mathbf w=-\mathbf v$. Hence $\min(m_+,m_-)\le\sqrt2|\mathbf r|$, with equality for pure states only at $\mathbf r=(\mathbf v+\mathbf w)/2\sqrt2$, the $H$-type states.
\end{proof}

Thus Spekkens contextuality and magic are inequivalent in this fragment. As Ref.~\cite{Spekkens2008} notes, negativity of one quasiprobability representation does not rule out a noncontextual model, and here the $\delta=\pm1$ model, a four-point Wigner function~\cite{WallmanBartlett2012}, is non-negative on $\mathrm{conv}(T_+)$ ($\delta=+1$) or $\mathrm{conv}(-T_+)$ ($\delta=-1$). The eight $T$-type magic states $\mathbf r=\mathbf v/\sqrt3$~\cite{BravyiKitaev2005} have $\varepsilon^*=0$~\cite{BozzettoHance2026ClassicalLimit} but $\varepsilon_{\rm stab}=1-1/\sqrt3\approx0.42$. Here $T$ is the order-three Clifford of Ref.~\cite{BravyiKitaev2005}. The state $(\ket0+e^{i\pi/4}\ket1)/\sqrt2$, obtained from $\ket+$ by the $\pi/8$ gate, is $H$-type, so it maximises $\varepsilon^*$. The pure states with $\varepsilon^*=0$ form eight spherical triangles, one about each $T$-type direction, with vertices at the three nearest Pauli eigenstates and sides on the circles of angular radius $\arccos(1/\sqrt3)$ about the three neighbouring $T$-type directions. Together they cover a fraction $4/\sqrt3-2\approx0.31$ of the Bloch sphere. Since $\varepsilon^*$ vanishes for $|\mathbf r|\le1/\sqrt2$, depolarising $\rho$ to visibility $1/\sqrt2$ or below makes the fragment simplex-embeddable whatever $\rho$ is, and the $H$-type states show that this bound is tight. If the noise also acts on the Pauli eigenstates, visibility $(2+\sqrt2)/4\approx0.85$ already suffices. For comparison, for the full qubit theory under depolarising noise of probability $p$, a preparation-noncontextual model exists if and only if $p\ge1/2$, and one that is also measurement noncontextual if and only if $p\ge2/3$~\cite{Marvian2020}, whereas $p=1-(2+\sqrt2)/4\approx0.15$ suffices for the stabiliser fragment with any $\rho$. We checked \cref{eq:epsqubit} against our own implementation of the linear program on $300$ random states (agreement to $4\times10^{-7}$), and for the global robustness we find $0.39<r/\varepsilon^*\le1/2$ over pure states, with $r=\varepsilon^*/2$ on the $H$-type states and infimum about $0.395$, so the range $0.40\le r/\varepsilon^*\le0.50$ of Appendix~\ref{app:lp} holds to two decimal places.

\begin{proposition}[Strange state]\label{prop:strange}
For $K=1$ qutrit, the fragment made of the twelve stabiliser states, the strange state $(\ket1-\ket2)/\sqrt2$ and the twelve stabiliser effects is simplex-embeddable, although the strange state has $W_{\min}=-1/3$.
\end{proposition}
\begin{proof}
Each single-qutrit stabiliser basis is a family of three parallel lines in phase space. The facets of the effect cone are the $81$ deterministic assignments of one outcome to each of the four stabiliser bases. A feasible point of the linear program of Ref.~\cite{Selby_2024} with rational entries (denominator $24$), symmetric under the $\mathrm{SL}(2,\Z_3)$ subgroup of the Clifford group that fixes the strange state, is given in the accompanying notebook and checked in exact arithmetic. In it the strange state is represented by weights $1/12$ and $1/24$ on eight assignments each, all $16$ of which avoid every line through the origin of phase space. Ref.~\cite{BozzettoHance2026ClassicalLimit} (Sec.~XI\,D and Table~I) also finds this fragment noncontextual, with models on $17$ and $18$ assignments.
\end{proof}

For one qutrit, as for one qubit, the set of states that can each be added to the stabiliser fragment without losing simplex-embeddability is therefore strictly larger than the set of states with non-negative Wigner function: on $60$ random pure states the state-only robustness from the linear program was below $9|W_{\min}|/(1+9|W_{\min}|)$, the value it would take if the region were the Wigner-positive set, for every Wigner-negative state. Ref.~\cite{BozzettoHance2026ClassicalLimit} (App.~D) finds the same for $20\,000$ Haar-random pure qutrit states. This is consistent with Ref.~\cite{Schmid2022StabSub}, where the uniqueness of the noncontextual model of the odd-dimensional stabiliser subtheory needs its Clifford transformations and diagram preservation, which fixes the number of ontic states to the dimension $d^2$ of the state space~\cite{Schmid2024Structure}. A prepare-and-measure fragment has neither. In the KS setting the single-qudit analogue is immediate, since each context is a single basis (fact 4 and Sec.~3.4 of Ref.~\cite{Delfosse2017}). \Cref{prop:strange} is the prepare-and-measure counterpart, in which the linear relations among the stabiliser states also constrain the model.

\subsubsection{Two or more logical qutrits}\label{sec:spekkens-multi}

\begin{proposition}[Two or more logical qutrits]\label{prop:multi}
For $K\ge2$ qutrits, the fragment made of all $K$-qutrit stabiliser states, a state $\rho$, and the projectors onto the stabiliser states grouped into the stabiliser bases is simplex-embeddable if and only if $W_\rho\ge0$. Hence, with $D=3^K$ and $W_{\min}=\min\{0,\min_uW_\rho(u)\}$,
\begin{equation}\varepsilon^*(\rho)=\frac{D^2|W_{\min}|}{1+D^2|W_{\min}|}.\label{eq:epsmulti}\end{equation}
\end{proposition}
\begin{proof}
If $W_\rho\ge0$, the Wigner representation of Gross is a simplex embedding: stabiliser states have non-negative Wigner functions, and the projector onto a stabiliser state is represented by the indicator function of an affine Lagrangian plane~\cite{Gross2006}. Conversely, take a simplex embedding, with ontic distributions $\mu_s$ for the states and response functions $\xi_E\in[0,1]$ for the effects, and let $\Lambda'$ be the support of $\mu_{\id/D}$. For every stabiliser basis $\{\ket{b_k}\}$, $\id/D=\sum_k\ketbra{b_k}{b_k}/D$, so by linearity $\Lambda'$ is the union of the supports of the $\mu_{b_k}$. Outcome $j$ has probability $\delta_{jk}$ on $\ket{b_k}$, so $\xi_{b_j}=\delta_{jk}$ on the support of $\mu_{b_k}$, and these supports are disjoint. Each $\lambda\in\Lambda'$ therefore fixes one outcome of every stabiliser basis, as in the outcome-determinism argument of Ref.~\cite{Spekkens2005}. Take the Heisenberg--Weyl operators $T_u=\omega^{-2^{-1}u_Z\cdot u_X}Z^{u_Z}X^{u_X}$, $u=(u_Z,u_X)\in\Z_3^{2K}$, with the phase convention of Ref.~\cite{Delfosse2017}, so that $T_{u+v}=T_uT_v$ whenever $[u,v]=0$. Each eigenprojector of $T_u$ is a sum of projectors of any stabiliser basis whose stabiliser group contains $T_u$, and by linearity its response function is the same for all such bases. So $\lambda$ assigns to each $T_u$ an eigenvalue $v_\lambda(u)$. If $[u,v]=0$, some stabiliser basis has $T_u$ and $T_v$ in its stabiliser group, and its joint eigenvalues give $v_\lambda(u+v)=v_\lambda(u)v_\lambda(v)$. The stabiliser states span the operator space and $\id/D$ is their uniform average, so $\id/D$ lies in the interior of the stabiliser polytope and $\id/D=p\rho+(1-p)\sigma$ with $p>0$ and $\sigma$ a mixture of stabiliser states, so the support of $\mu_\rho$ lies in $\Lambda'$~\cite{Schmid2022StabSub}. Then $\mathrm{Tr}(T_u\rho)=\sum_{\lambda\in\Lambda'}\mu_\rho(\lambda)v_\lambda(u)$ for every $u$. This is a set of noncontextual value assignments for $\rho$ in the sense of Ref.~\cite{Delfosse2017}, and for $K\ge2$ it implies $W_\rho\ge0$ (Proposition~1 of Ref.~\cite{Delfosse2017}). Finally, $W_{\id/D}=1/D^2$ at every point, so $(1-\varepsilon)\rho+\varepsilon\id/D$ has a non-negative Wigner function if and only if $\varepsilon$ is at least the value in \cref{eq:epsmulti}.
\end{proof}

The ``only if'' part follows the route sketched in Appendix~D of Ref.~\cite{Schmid2022StabSub}, with the $n\ge2$ result of Ref.~\cite{Delfosse2017} in place of the single-qudit one with an ancilla~\cite{Howard2014}. For $K=1$ this step fails: a single qutrit has $81$ value assignments but only $9$ phase-space points~\cite{BozzettoHance2026ClassicalLimit} (both counts are checked in the accompanying notebook), and \cref{prop:strange} shows that the fragment can then be simplex-embeddable for a Wigner-negative state. For $K\ge2$ logical qutrits, Spekkens contextuality of the logical state thus has the same zero set as the mana and the contextual fraction (\cref{prop:zeroset}), and $\varepsilon^*$ depends on $\rho$ only through $W_{\min}$, so its circuit average follows from the same Wigner functions as the mana. \Cref{eq:epsmulti} has the form of the depolarising thresholds of Ref.~\cite{vanDamHoward2011} (Lemma~2 there). For example, $\frac{1}{\sqrt2}(\ket1-\ket2)\otimes\ket0$ has $W_{\min}=-1/9$ and $\varepsilon^*=9/10$, although the strange state alone has $\varepsilon^*=0$: the two-qutrit fragment contains entangled bases that the one-qutrit fragment lacks. More generally, appending a pure stabiliser qutrit to any state divides $W_{\min}$ by $3$ and multiplies $D^2$ by $9$, so $D^2|W_{\min}|$ triples and $\varepsilon^*$ moves towards $1$. Unlike the mana, $\varepsilon^*$ is therefore not additive over logical qutrits.

\section{First result: Wigner negativity}
We simulate the exact state vector of $N$ qutrits, apply the encoder, the error layer for $21$ angles $\alpha\in[0,2\pi/3]$ ($11$ for $N=12$ and for $(N,K)=(10,5)$)
, and the decoder, and for every syndrome evaluate the Wigner function of the logical state. All syndromes are included, weighted by their probability, so the only sampling error is the circuit average. The numbers of circuits are given in the figure captions. We found the same structure found for magic in Ref.~\cite{Niroula_2024}.

\subsection{Vanishing rate}\label{sec:vanishing}
\begin{figure}[t]
\includegraphics[width=0.45\textwidth]{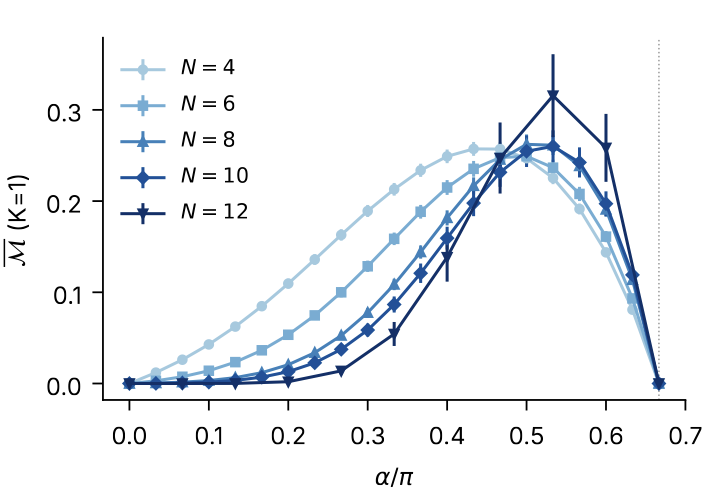}
\caption{Vanishing rate ($K=1$) with $N=4,6,8,10,12$ qutrits (400, 300, 200, 80, 8 circuits). Circuit-averaged mana against the error angle; the dotted line marks the Clifford point $\alpha=2\pi/3$. Error bars are standard errors over circuits.}\label{fig:k1}
\end{figure}

For $K=1$ the averaged mana $\overline\Mana(\alpha)$ is shown in \cref{fig:k1}. It vanishes at both Clifford points ($\alpha=0$ and $\alpha=2\pi/3$) and has a single peak. The peak moves towards $\alpha=2\pi/3$ as $N$ grows. Plotted against $x=(2\pi/3-\alpha)\sqrt N$, the curves for $N=4$--$12$ lie on one function within statistical error. Over $N=4$--$12$ the peak position is $x^*=1.37(2)$, constant within error, which is the $1/\sqrt N$ law of Ref.~\cite{Niroula_2024} transposed to qutrits. For qubits with $Z$-rotation errors and a fixed number of logical qubits, Ref.~\cite{Feng2026} argues that no magic phase persists at fixed $\alpha$ as $N\to\infty$. A peak that approaches the Clifford point as $1/\sqrt N$ is consistent with this. The peak height is $0.25$--$0.26$ for $N=4$--$10$, independent of $N$. Near the Clifford point $\overline\Mana\simeq0.43\,x$, linear in $x$, whereas the SSRE of Ref.~\cite{Niroula_2024} grows as $N(\pi/2-\alpha)^2/4$, quadratically in the corresponding scaling variable. The collapse is good for $x\lesssim3$; for larger $x$ the curves at these sizes are also affected by the other Clifford point $\alpha=0$ (\cref{fig:k1}). 

\subsection{Constant rate}\label{sec:rate}
\begin{figure*}
\includegraphics[width=0.9\textwidth]{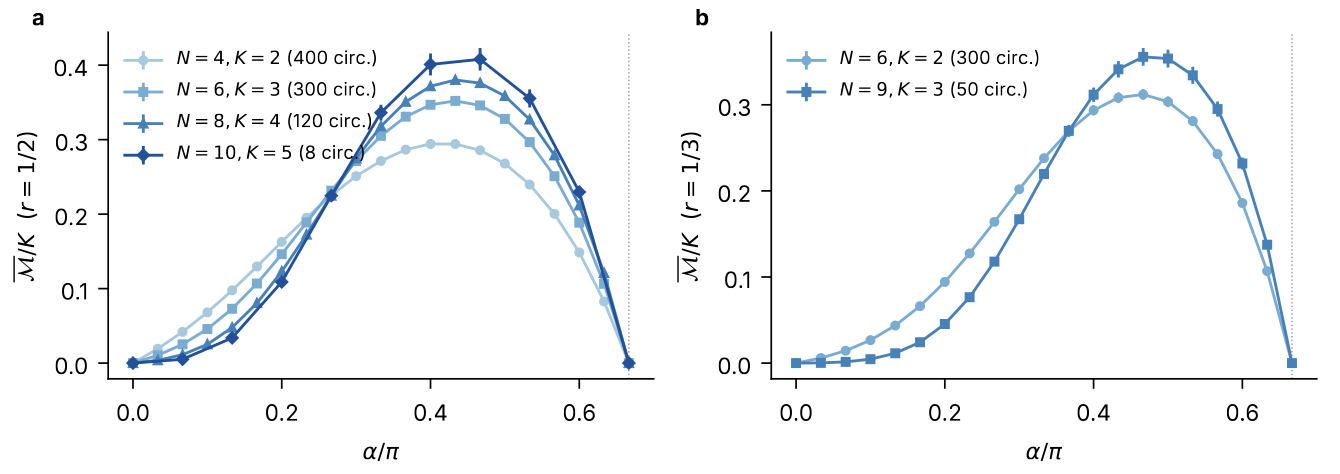}
\caption{Constant rate. Mana density $\overline\Mana/K$ against $\alpha/\pi$. (a) $r=1/2$ with $(N,K)=(4,2),(6,3),(8,4),(10,5)$ and $400,300,120,8$ circuits. (b) $r=1/3$ with $(6,2),(9,3)$ and $300,50$ circuits. Dotted lines: Clifford point $2\pi/3$.}\label{fig:rate}
\end{figure*}

For constant rate we show the mana density $\overline\Mana/K$ at $r=1/2$ ($N=4,6,8,10$ with $K=2,3,4,5$) and at $r=1/3$ ($N=6,9$ with $K=2,3$) in \cref{fig:rate}. At small error angle the density decreases with $N$; above a crossing it increases with $N$. At $r=1/2$ the crossings of consecutive sizes are at $\alpha/\pi=0.29\pm0.01$ ($N=4\to6$), $0.27\pm0.01$ ($6\to8$) and $0.28\pm0.02$ ($8\to10$, where $N=10$ has only 8 circuits), with no resolved drift. A scaling collapse $\overline\Mana/K^\gamma=F\bigl((\alpha/\pi-\alpha_c/\pi)N^{1/\nu}\bigr)$ with $\gamma=1$ over $N=4$--$10$ gives $\alpha_c/\pi\approx0.26$ (range $0.255$--$0.275$ at twice the minimal collapse cost) and $\nu\approx1.25$ (range $1.2$--$1.5$). At $r=1/3$ the single crossing of $(N,K)=(6,2)$ and $(9,3)$ is at $\alpha/\pi=0.37$, i.e.\ at larger angle for smaller rate, as expected since fewer logical qutrits means more syndrome measurements.

For comparison, Ref.~\cite{Niroula_2024} reports for qubits at $r=1/2$ the SSRE values $\alpha_c/\pi=0.27(1)$, $\nu=1.15(4)$ with $\gamma=1.20(8)$ fitted, and $\alpha_c/\pi=0.240(2)$, $\nu=0.96(5)$ with $\gamma$ constrained to $[0,1]$, where the fit gives $\gamma=1$, and, for the conditional entropy, $\alpha_c/\pi=0.304(2)$, $\nu=2.9(2)$. Our numbers are not directly comparable (qutrits, mana, and sizes up to $N=10$ against $N\le14$ for the SSRE and $N\le24$ for the conditional entropy), and the exponent in particular is poorly constrained by four small sizes. We therefore read Fig. \ref{fig:rate} as evidence that the mana density shows the same two-regime structure (suppression below a crossing, concentration above it), not as a determination of critical exponents. With Born-rule syndrome sampling at $r=1/2$, Ref.~\cite{Sierant2026} finds that the crossing of the logical fidelity drifts with $N$ up to $N\approx20$ before settling, and locates the SSRE critical point by extrapolating crossings in $1/N$, so crossings at $N\le10$ should not be read as estimates of $\alpha_c$.

\section{Second result: KS contextuality}\label{sec:ksresult}
Since the size of the linear program for the contextual fraction of \cref{eq:cf} grows quickly with $K$ (\cref{sec:ks}), we study $K=2$ with growing $N$, a vanishing-rate regime. For each circuit and angle we sample $20$ syndromes from $p_C(s;\alpha)$ and solve the linear program for the logical two-qutrit state; the mean is an unbiased estimate of $\sum_sp_C(s;\alpha)\CF(\psi_{C,s})$. We use $80,60,40$ circuits for $N=4,6,8$ and $11$ angles.
\begin{figure}[t]
\includegraphics[width=0.45\textwidth]{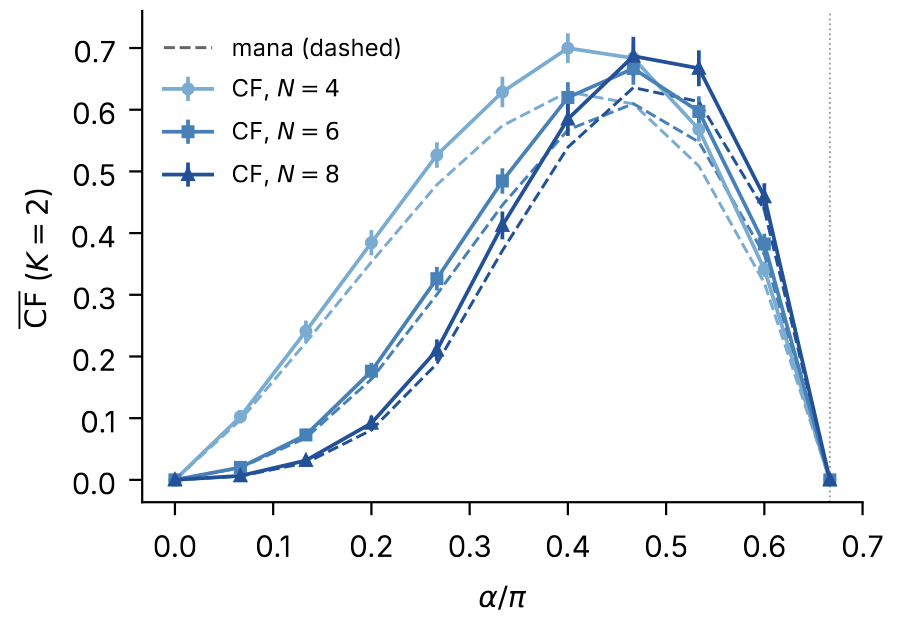}
\caption{Contextual fraction (symbols) and mana (dashed) of the two-qutrit logical state against $\alpha/\pi$, for $N=4,6,8$. Standard errors on $\overline{\CF}$ are at most $0.03$.}\label{fig:cf}
\end{figure}
The contextual fraction vanishes at the Clifford points, as it must (\cref{prop:zeroset}), is at least the mana at every angle within error, and tracks it closely (\cref{fig:cf}). The two are not ordered in general: $\CF\le1$, whereas the product of two strange states has $\CF=1$ and $\Mana=2\log(5/3)\approx1.02$, and about $40\%$ of Haar-random two-qutrit pure states have $\CF<\Mana$. It decreases with $N$ at small angle and its maximum moves towards the Clifford point $\alpha=2\pi/3$: $\max\overline{\CF}=0.70$ at $\alpha/\pi=0.40$ for $N=4$, $0.67$ at $0.47$ for $N=6$ and $0.69$ at $0.47$ for $N=8$, i.e.\ a height of about $0.7$ independent of $N$, as for the mana. With only eleven angles the peak position is resolved too coarsely for a quantitative $1/\sqrt N$ test. So in the KS sense the same structure appears, with a baseline that is exactly zero at the Clifford points.

\section{Third result: Spekkens contextuality}\label{sec:spekkens}
For $K=1$ the stabiliser fragments of one qubit and one qutrit are simplex-embeddable; for two qubits even Peres--Mermin sub-fragments are not (Appendix~\ref{app:lp}), so we restrict to $K=1$.

\begin{figure*}
\includegraphics[width=0.9\textwidth]{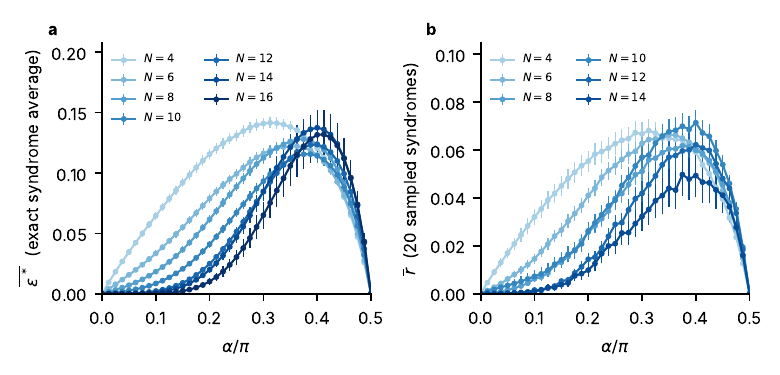}
\caption[]{Single qubit results of robustness of Spekkens contextuality. \textbf{a} robustness of simplex embeddability $\bar\varepsilon^*$ as function of $\alpha/\pi$, for different number of total qubits, averaged over all syndromes, \textbf{b} global robustness of simplex embeddability $\bar r$ as function of $\alpha/\pi$, for different number of total qubits, from $20$ syndromes sampled per circuit and angle. The number of circuits ranges from 30 to 150, depending on the number of qubits (from 150 circuits for the case where $N=4$ to 30 for the case where $N=16$ for $\bar\varepsilon^*$ and $N=14$ for $\bar{r}$).
}\label{fig:reps-qubit}
\end{figure*}

For the single qubit we can see the results in Fig. \ref{fig:reps-qubit}.

Both quantities vanish at $\alpha=0$ and at the Clifford point $\alpha/\pi = 0.5$, and both have one peak. The peak moves towards the Clifford point as $N$ grows.
The $r$ peak is about 0.05 to 0.07, roughly half the $\varepsilon^*$ peak of 0.12 to 0.14. By \cref{prop:qubit}, $\varepsilon^*$ never exceeds the white-noise robustness of magic $\varepsilon_{\rm stab}$. In our simulations of the same ensemble (all syndromes), at the peak of $\overline{\varepsilon_{\rm stab}}$ the ratio $\overline{\varepsilon^*}/\overline{\varepsilon_{\rm stab}}$ is $0.95(1)$ for $N=4$, $0.87(2)$ for $N=8$ and $0.85(3)$ for $N=12$ ($600$, $300$ and $120$ circuits), and the syndrome weight on magic but noncontextual logical states is at most $0.017(4)$, $0.042(8)$ and $0.046(11)$. The two peaks lie at the same angle within our grid. So for one logical qubit the peak in Spekkens contextuality sits where the magic peak does, but is lower, increasingly so at larger $N$.

\begin{figure*}
\includegraphics[width=0.9\textwidth]{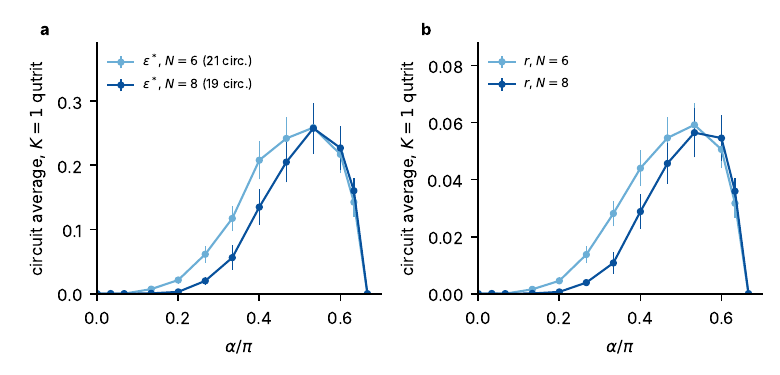}
\caption[]{Single qutrit results of robustness of Spekkens contextuality. \textbf{a} robustness of simplex embeddability $\bar\varepsilon^*$ as function of $\alpha/\pi$, for different number of total qutrits, \textbf{b} global robustness of simplex embeddability $\bar r$ as function of $\alpha/\pi$, for different number of total qutrits. The number of circuits simulated is equal to 21 for $N=6$ and 19 for $N=8$. }\label{fig:reps-qutrit}
\end{figure*}

For the single qutrit we can see the results in Fig. \ref{fig:reps-qutrit}.

Both quantities vanish at $\alpha=0$, and both have one peak, near $\alpha/\pi\simeq0.53$ for both sizes. The $\varepsilon^*$ peak is about 0.26, while the $r$ peak is about 0.056. Our own implementation of the $81$-vertex linear program gives $\overline{\varepsilon^*}=0.26\pm0.06$ at $\alpha/\pi=0.53$ for $N=6$ ($10$ circuits, $8$ sampled syndromes each), in agreement with \cref{fig:reps-qutrit}; a noncontextual set equal to the Wigner-positive states would give $0.35\pm0.08$ on the same samples, larger for every Wigner-negative sample (\cref{prop:strange}).

\begin{figure*}
\includegraphics[width=\textwidth]{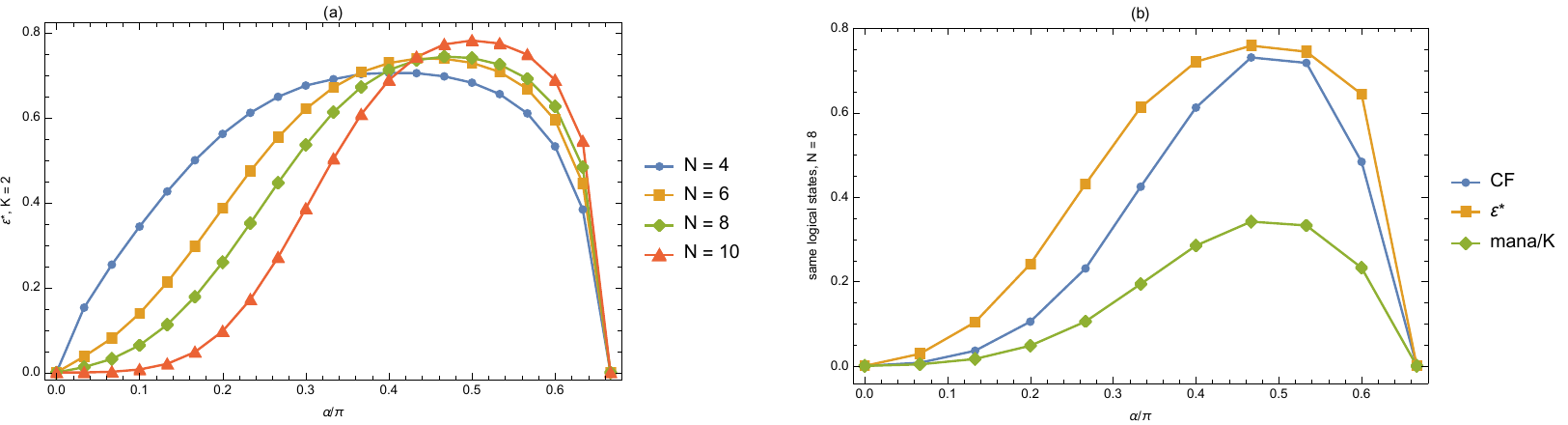}
\caption{Spekkens contextuality of two logical qutrits at vanishing rate. (a) Circuit-averaged noise robustness $\overline{\varepsilon^*}$, from \cref{eq:epsmulti}, for $N=4,6,8,10$ ($400$, $300$, $80$, $16$ circuits); standard errors are at most $0.022$. (b) Contextual fraction, $\varepsilon^*$ and mana per logical qutrit on the same logical states, sampled with $20$ syndromes per circuit and angle, for $N=8$ ($40$ circuits).}\label{fig:spk-k2}
\end{figure*}

\begin{figure*}
\includegraphics[width=\textwidth]{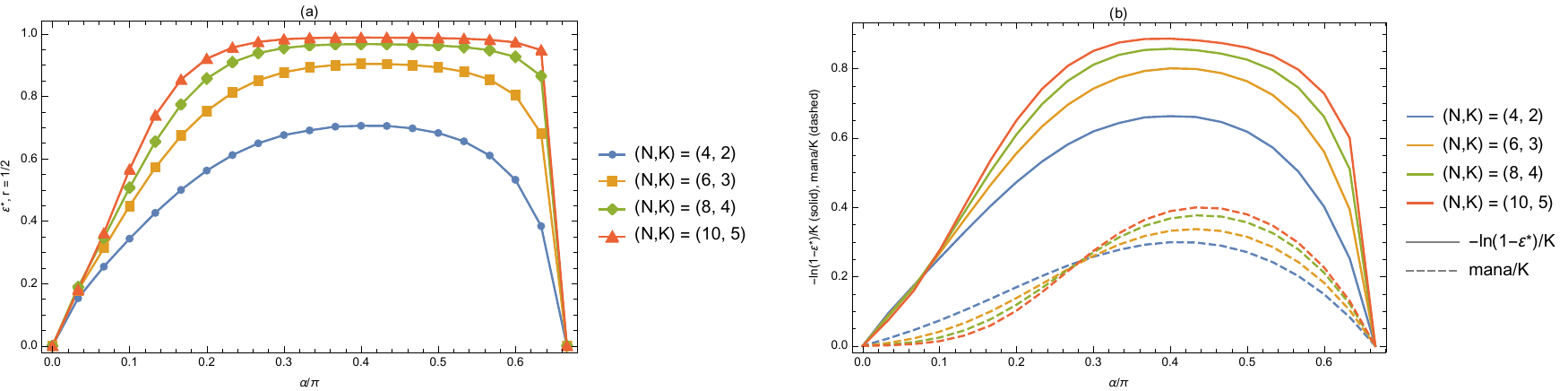}
\caption{Spekkens contextuality at constant rate $r=1/2$, for $(N,K)=(4,2),(6,3),(8,4),(10,5)$ ($400$, $300$, $120$, $24$ circuits). (a) Circuit-averaged noise robustness $\overline{\varepsilon^*}$; standard errors are at most $0.026$, except for $(10,5)$ at $\alpha/\pi\le0.13$, where they reach $0.059$. (b) $-\ln(1-\varepsilon^*)/K$ (solid) and the mana density (dashed).}\label{fig:spk-rate}
\end{figure*}

\emph{Two or more logical qutrits.} For $K\ge2$, \cref{eq:epsmulti} gives $\varepsilon^*$ of each logical state from its Wigner function, so we computed $\overline{\varepsilon^*}$ for the circuits of the same model, with exact syndrome weights (\cref{fig:spk-k2,fig:spk-rate}). At vanishing rate with $K=2$ it behaves like the mana. It vanishes at the Clifford points, and its peak, $0.705\pm0.009$, $0.739\pm0.008$, $0.744\pm0.017$ and $0.782\pm0.004$ for $N=4,6,8,10$, moves from $\alpha/\pi=0.40$ to $0.50$, as the peak of the mana does. On the same sampled logical states, $\overline{\varepsilon^*}\ge\overline{\CF}\ge\overline\Mana/K$ at every angle for $N=4,6,8$.

At constant rate the two behave differently. For $r=1/2$, $\overline{\varepsilon^*}$ increases with $N$ by more than two standard errors at $18$, $17$ and $16$ of the $19$ interior angles from $(N,K)=(4,2)$ to $(6,3)$, from $(6,3)$ to $(8,4)$ and from $(8,4)$ to $(10,5)$, and it decreases significantly at none. Its peak, at $\alpha/\pi=0.40$, rises from $0.705$ to $0.903$, $0.967$ and $0.988$. Over the same runs the mana density keeps its crossings, at $\alpha/\pi=0.29\pm0.01$, $0.27\pm0.01$ and $0.28\pm0.02$.
For $r=1/3$ the peak of $\overline{\varepsilon^*}$ rises from $0.739$ at $(6,2)$ to $0.913$ at $(9,3)$. This is the non-additivity noted after \cref{prop:multi}. As $K$ grows, $\varepsilon^*$ approaches $1$, so the noise robustness of Spekkens contextuality does not show the constant-rate transition, and the trend in these runs is towards a step between the Clifford points and all other angles. The finite form $-\ln(1-\varepsilon^*)/K=\ln(1+D^2|W_{\min}|)/K$ does not show it either: it also grows with $N$ above $\alpha/\pi\approx0.1$, and its crossings, near $\alpha/\pi\approx0.07$--$0.10$, are poorly determined.

\section{Hardware experiment}\label{sec:hardware}

\emph{Protocol.} We ran the $K=1$ model of \cref{sec:setup-qubit} on the five-qubit superconducting processor (IQM Spark, star connectivity) of Politecnico of Turin with $N=5$: one logical qubit and four syndrome qubits. Each circuit consists of $d=3$ layers of random single-qubit Cliffords followed by CZ gates on a random matching of the available edges (the star admits one CZ per layer), the coherent error $R_z(\alpha)$ on every qubit (a virtual gate on this platform), and the inverse of the encoder. The four syndrome qubits are measured in the $Z$ basis and the logical qubit in the $X$, $Y$ and $Z$ bases. We used $8$ random circuits, $9$ angles $\alpha/\pi\in\{0,0.1,0.2,0.25,0.3,0.35,0.4,0.45,0.5\}$, $3$ bases and $4000$ shots per setting, with $\langle\text{CZ}\rangle=5.694$ two-qubit gates per transpiled circuit. 
For each syndrome outcome $s$ we reconstruct the Bloch vector $\mathbf r_s$ of the logical qubit from the three bases, clipping to the Bloch ball when shot noise places the estimate outside it, and discard outcomes with fewer than $15$ counts in any basis. For each $\mathbf r_s$ we compute the state-only robustness $\varepsilon^*$ and the global robustness $r$ (the eight-vertex form of the linear program), and average over outcomes with the measured syndrome probabilities, then over circuits. As a reference we compute the same quantities exactly and noiselessly for the same circuits. Error bars are the standard error over the $8$ circuits; they are dominated by circuit-to-circuit variation, which is also present in the exact curve.

\begin{figure*}
\includegraphics[width=0.9\textwidth]{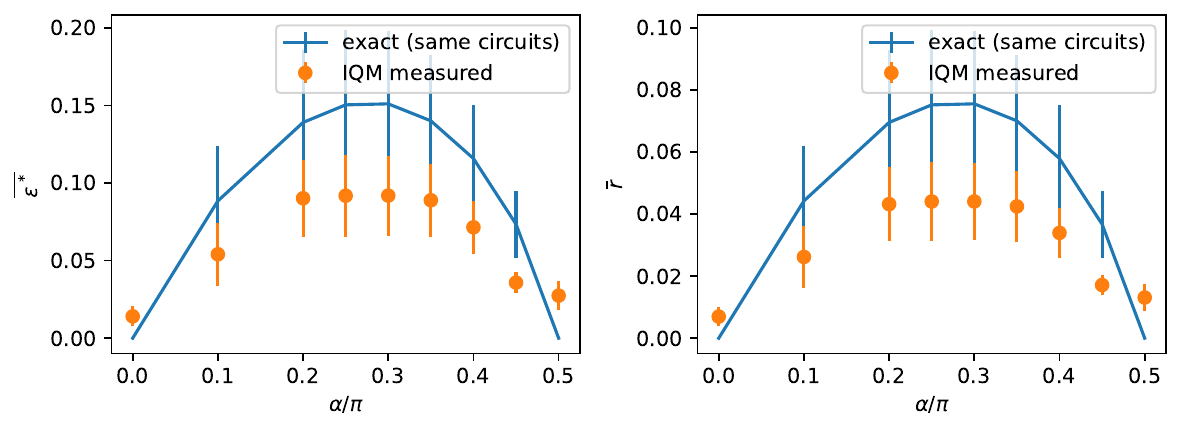}
\caption{Circuit-averaged $\overline{\varepsilon^*}$ (left) and $\overline r$ (right) of the logical qubit against $\alpha/\pi$ for $N=5$, $d=3$. Blue: exact noiseless values for the same $8$ circuits. Orange: measured. Error bars of exact and real-hardware measured data: standard error over circuits. 
}\label{fig:hw-iqm}
\end{figure*}

\begin{table}[t]
\begin{ruledtabular}
\begin{tabular}{c cc cc}
$\alpha/\pi$ & $\overline{\varepsilon^*}_{\rm hw}$ & $\overline{\varepsilon^*}_{\rm exact}$ & $\overline r_{\rm hw}$ & $\overline r_{\rm exact}$\\\hline
0.00 & 0.014 & 0.000 & 0.007 & 0.000\\
0.10 & 0.054 & 0.088 & 0.026 & 0.044\\
0.20 & 0.090 & 0.139 & 0.043 & 0.070\\
0.25 & 0.092 & 0.150 & 0.044 & 0.075\\
0.30 & 0.092 & 0.151 & 0.044 & 0.076\\
0.35 & 0.089 & 0.140 & 0.042 & 0.070\\
0.40 & 0.072 & 0.116 & 0.034 & 0.058\\
0.45 & 0.036 & 0.073 & 0.017 & 0.037\\
0.50 & 0.027 & 0.000 & 0.013 & 0.000
\end{tabular}
\end{ruledtabular}
\caption{Values plotted in \cref{fig:hw-iqm}.}\label{tab:hw-iqm}
\end{table}

\emph{Results.} The measured curves have the shape of the exact ones (\cref{fig:hw-iqm}, \cref{tab:hw-iqm}). Both $\overline{\varepsilon^*}$ and $\overline r$ are small near the Clifford points, rise to a single maximum at $\alpha/\pi\simeq0.25$--$0.30$, which is the position of the exact maximum, and decrease again towards $\alpha/\pi=0.5$. The measured maxima, $0.092$ for $\overline{\varepsilon^*}$ and $0.044$ for $\overline r$, are about $60-65\%$ of the exact ones ($0.151$ and $0.076$) for both quantities, and the measured curve lies below the exact one at every interior angle. The ratio $\overline r/\overline{\varepsilon^*}$ is $\simeq0.48$ on hardware and $\simeq0.50$ in the exact calculation, as expected from the pointwise relation $0.40\le r/\varepsilon^*\le0.50$ found for qubits.

\emph{Interpretation.} A uniform shrinking of the logical Bloch vectors by a factor $v$ lowers $\varepsilon^*$ and $r$ more than proportionally, because both vanish on the set of vectors inside the noncontextual region (for $\varepsilon^*$, the two tetrahedra of \cref{prop:qubit}). The strongest suppression is at $\alpha/\pi=0.45$ (about $1/2$ of the exact value). At the Clifford points $\alpha/\pi=0$ and $0.5$ the exact values are zero, and so would be those of any state that is a mixture of stabiliser states, in particular of a depolarised stabiliser state. The residual values $0.014$ and $0.027$ for $\overline{\varepsilon^*}$ and $0.007$ and $0.013$ for $\overline r$ (about $9\%$ and $17-18\%$ of the exact maximum) must therefore come from other sources: finite-shot bias, since a pure stabiliser vector lies on the boundary of the noncontextual region and statistical scatter moves it outside, readout asymmetry, and coherent errors of the processor. We did not separate these contributions. This is a practical advantage of the two robustness measures over the stabiliser R\'enyi entropy, which is non-zero for a depolarised stabiliser state.

\emph{Limits of the test.} Five qubits with a star connectivity allow a single logical qubit and a weak scrambler, so the experiment tests the existence, position and height of the peak and its vanishing at the Clifford points, but not the $1/\sqrt N$ scaling of its position. The statistics are small ($8$ circuits), the comparison with the exact curve for the same circuits is systematic rather than statistical, and the circuit ensemble (star connectivity, $d=3$, CZ gates) differs from the all-to-all ensemble used in the numerics, so the exact reference must be computed for the executed circuits, as done here. The stabiliser states and Pauli effects of the fragment are taken to be ideal and only the logical state is measured, so the experiment estimates $\varepsilon^*$ and $r$ within quantum theory. A direct test of noncontextuality would also have to avoid assuming noiseless measurements and exact operational equivalences~\cite{Mazurek2016}.

\begin{figure*}
\includegraphics[width=0.9\textwidth]{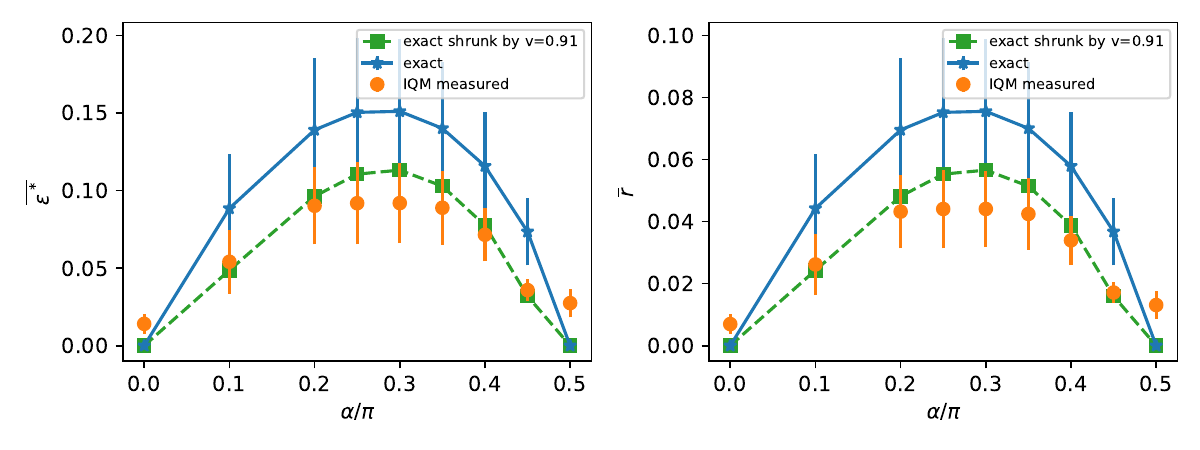}
\caption[]{Circuit-averaged $\overline{\varepsilon^*}$ (left) and $\overline r$ (right) of the logical qubit against $\alpha/\pi$ for $N=5$, $d=3$. Blue: exact noiseless values for the same $8$ circuits. Orange: measured. Error bars of exact and real-hardware measured data: standard error over circuits. Green dashed: exact values for Bloch vectors shrunk by $v=0.91$.}
\label{fig:visibility}
\end{figure*}

\emph{Visibility.}
Visibility is 0.91. Shrinking the exact Bloch vectors by $v=0.91$ leaves about $75\%$ of the exact peak (\cref{fig:visibility}). Visibility has been calculated through the following formula
\begin{equation}
\begin{aligned}
v&=\frac{\sum p\,\sum_{P\in\{X,Y,Z\}}\langle P\rangle_{\rm hw}\langle P\rangle_{\rm ex}}
       {\sum p\,\sum_{P\in\{X,Y,Z\}}\langle P\rangle_{\rm ex}^{2}}= \\
&=\; \frac{\displaystyle\sum_{c}\sum_{\alpha}\sum_{s}\, p_{c,\alpha}(s)\;\mathbf r^{\rm hw}_{c,\alpha,s}\cdot\mathbf r^{\rm ex}_{c,\alpha,s}}
             {\displaystyle\sum_{c}\sum_{\alpha}\sum_{s}\, p_{c,\alpha}(s)\;\bigl|\mathbf r^{\rm ex}_{c,\alpha,s}\bigr|^{2}}      
\label{eq:visibility}
\end{aligned}
\end{equation}
where $c$ is the index of the random circuit, $\alpha$ the error angle, $s$ the syndrome outcome, $p_{c,\alpha}(s)$ the exact noiseless probability of getting syndrome $s$ for circuit $c$ at angle $\alpha$, $\textbf{r}^{\text{ex}}$ the exact noiseless Bloch vector of the logical qubit and $\textbf{r}^{\text{hw}}$ the same Bloch vector reconstructed from the measured counts in the X, Y and Z bases.

We define the logical visibility $v$ (\cref{eq:visibility}) as the least-squares factor mapping the exact Bloch vectors onto the measured ones. It absorbs the average effect of every noise source on the executed circuits, and for our data $v = 0.91$. At the time of the run (14/11/2026, 10.05 am) the device's published calibration~\cite{QuantumComputerPoliTO} gave readout fidelities from $94.18\%$ to $99.04\%$, $T_1$ from 26 to 44 $\mu s$, $T_2$  8 to 15 $\mu s$ and PRX and CZ gate fidelities above $99.26\%$.  Shrinking the exact Bloch vectors by $v = 0.91$ leaves about $75\%$ of the exact peak~\cref{fig:visibility}, against about $60\%$ measured. From $\alpha/\pi = 0.2$ to $0.4$ the measured $\varepsilon^*$ and $r$ lie below the shrunk curves at every angle, which suggests that uniform depolarisation accounts for most of the suppression but not all of it. 

\begin{figure*}
\includegraphics[width=0.9\textwidth]{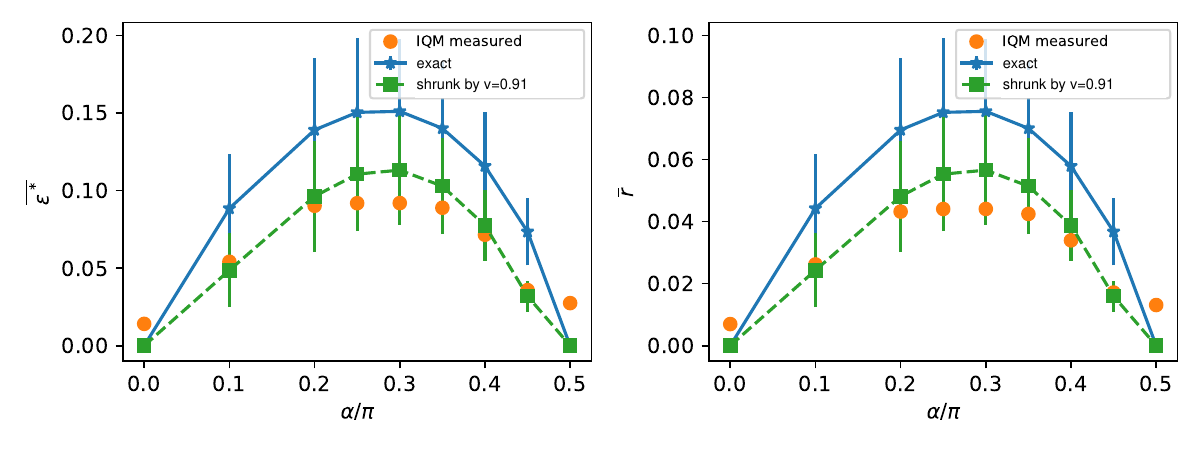}
\caption{Circuit-averaged $\overline{\varepsilon^*}$ (left) and $\overline r$ (right) of the logical qubit against $\alpha/\pi$ for $N=5$, $d=3$. Blue: exact noiseless values for the same $8$ circuits. Orange: measured. Error bars of exact and exact shrunk data: standard error over circuits. Green dashed: exact values for Bloch vectors shrunk by $v=0.91$.}\label{fig:errorbars}
\end{figure*}

\section{Discussion}\label{sec:discussion}
For qutrits, KS contextuality with respect to stabiliser measurements has the same zero set as Wigner negativity (\cref{prop:zeroset}), and the contextual fraction of two logical qutrits tracks the mana closely. In this sense the vanishing-rate magic peak of Ref.~\cite{Niroula_2024} has a KS-contextuality counterpart; whether the constant-rate transition has one is open, since we computed the contextual fraction only for $K=2$. For Spekkens contextuality the question has an answer for $K\ge2$: by \cref{prop:multi} the noise robustness follows from the Wigner function, and at constant rate it approaches $1$ at all non-Clifford angles instead of showing the transition (\cref{fig:spk-rate}). In the prepare-and-measure fragments used for Spekkens contextuality the two notions are not equivalent: for one qubit the robustness vanishes on all $T$-type magic states (\cref{prop:qubit} and Ref.~\cite{BozzettoHance2026ClassicalLimit}), and for one qutrit on the strange state (\cref{prop:strange}). For two or more logical qutrits they are equivalent again (\cref{prop:multi}), so the separation is a single-qudit effect. For one qubit the circuit averages still follow the magic peak, because only a few per cent of the syndrome weight falls on magic but noncontextual states, but the gap grows with $N$ over the sizes we reach. On hardware the single-qubit peak survives at about $60\%$ of its noiseless height (\cref{sec:hardware}); by \cref{prop:qubit}, any visibility below $1/\sqrt2$ would remove it.

Including the coherent error as a transformation changes the picture. In odd dimensions the stabiliser subtheory, with its Clifford transformations, has a unique diagram-preserving noncontextual model, the Wigner representation of Gross~\cite{Gross2006}, and adding any state that promotes it to universal computation, such as the non-stabiliser pure state $U(\alpha)\ket{+}$ for $\alpha\notin(2\pi/3)\Z$, leaves no convex-linear, diagram-preserving noncontextual model~\cite{Schmid2022StabSub,BozzettoHance2026ClassicalLimit}. A scenario that contains the stabiliser subtheory together with the coherent error is therefore contextual in this sense for every non-Clifford $\alpha$, whatever $N$, $K$ and the syndrome. The noise threshold depends on how transformations are represented: for the depolarised strange state with the Clifford gates as transformations, contextuality persists up to depolarising probability $3/4$ in the convex-linear reading and $3/5$ in a weaker class-representative reading~\cite{BozzettoHance2026ClassicalLimit}. The value $3/4$ is also the depolarising probability at which the strange state enters the stabiliser polytope, the largest for any qutrit state~\cite{vanDamHoward2011}. For qubits the stabiliser subtheory with its Clifford transformations is already contextual in the convex-linear reading~\cite{Lillystone2019,Schmid2022StabSub}, so there the zero baseline is lost even at the Clifford points. A transition in contextuality is thus a property of the logical states, not of the process that produces them.

For two or more logical qubits, the full Pauli scenario has no noncontextual baseline (\cref{sec:setup-qubit}), and the Peres--Mermin sub-fragments of Appendix~\ref{app:lp} are already contextual, so a qubit order parameter for $K\ge2$ needs a smaller fragment. Restricted qubit settings without state-independent contextuality exist, such as rebits with CSS states~\cite{Delfosse2015Rebits} and qubit schemes whose free measurements preserve Wigner non-negativity~\cite{Raussendorf2017}, but there contextuality implies Wigner negativity and not conversely~\cite{Delfosse2015Rebits,Raussendorf2017}, so the zero sets need not coincide as in \cref{prop:zeroset}. The phase-space method for qubits based on closed noncontextual sets~\cite{Raussendorf2020} represents all stabiliser mixtures non-negatively, and the $\Lambda$-polytope model~\cite{Zurel2020} every state, but the first is contextual by the usual definitions and the second is preparation contextual~\cite{Raussendorf2020,Zurel2020}, so they do not supply one directly. Since \cref{eq:cf} sees $\rho$ only through sums of $W_\rho$ over affine Lagrangian planes, two-sided bounds between $\CF$ and the sum negativity would carry the constant-rate data over to KS contextuality.

\textit{Acknowledgments---}
JRH acknowledges support from a Royal Society Research Grant (RG/R1/251590), from an EPSRC Mathematical Sciences Small Grant (UKRI3647), and from their EPSRC Quantum Technologies Career Acceleration Fellowship (UKRI1217).

\textit{Data and Code Availability---} A zip file containing all the codes used in this paper is openly available on Zenodo~\cite{BozzettoHance2026Notebook}.

\bibliographystyle{unsrturl}
\bibliography{ref.bib}

\appendix
\section{Linear-program checks of the fragments}\label{app:lp}
We used the code released with Ref.~\cite{Selby_2024} and we found that the global robustness $r$ of \cref{eq:rglobal} vanishes on exactly the same states and satisfies $0.40\le r/\varepsilon^*\le0.50$ on $1500$ random pure qubit states simulated. This shows that global depolarisation makes the test about twice as permissive.

For $K=1$ the stabiliser fragment is embeddable for both qubits and qutrits: the qubit fragment (six states,  $8$ vertices) and the qutrit fragment (twelve states, $81$ vertices) both give residual (the number the linear program minimises to decide embeddability) equal to 0; here the vertices are the deterministic assignments of outcomes to the $3$ (qubit) or $4$ (qutrit) stabiliser bases, the candidate ontic states of an embedding. 
For two qubits it is not. A sub-fragment made of the Pauli observables of the Peres--Mermin square already fails: one context (four effects) is embeddable, but the three rows together (42 states, 12 effects, $\dim V=10$, $64$ vertices) give a residual of $0.8$, and so does the full square (24 effects, 120 vertices). The robustness of \cref{eq:rglobal}, computed with the code of Ref.~\cite{Selby_2024} on the same sub-fragments (all $60$ two-qubit stabiliser states; the Pauli effects of the contexts listed), is $r=0$ for one context and $r=0.4444$ ($=4/9$ to the printed precision) for the three rows, for the rows plus a column, and for the full square.

\end{document}